\documentclass[11pt]{article}
\usepackage{moriond,epsfig}
\usepackage{amsmath}
\usepackage{graphicx}
\usepackage{subfigure}

\begin{document}
\vspace*{4cm}
\title{DIRECT CP VIOLATION IN D-MESON DECAYS}

\author{ JOACHIM~BROD }

\address{Department of Physics, University of Cincinnati, Cincinnati,
  OH 45221, USA}

\maketitle \abstracts{ Recently, the LHCb and CDF collaborations
  reported a surprisingly large difference between the direct CP
  asymmetries, $\Delta {\cal A}_{CP}$, in the $D^0\to K^+K^-$ and
  $D^0\to \pi^+ \pi^-$ decay modes. An interesting question is whether
  this measurement can be explained within the standard model.  In
  this review, I would like to convey two messages: First, large
  penguin contractions can plausibly account for this measurement and
  lead to a consistent picture, also explaining the difference between
  the decay rates of the two modes. Second, ``new physics''
  contributions are by no means excluded; viable models exist and can
  possibly be tested. }

\section{Introduction}

The $D^0\to K^+K^-$ and $D^0\to \pi^+ \pi^-$ decays are induced by the
weak interaction via an exchange of a virtual $W$ boson and suppressed
by a single power of the Cabibbo angle. Direct $CP$ violation in
singly Cabibbo-suppressed (SCS) $D$-meson decays is sensitive to
contributions of new physics in the up-quark sector, since it is
expected to be small in the standard model: the penguin amplitudes
necessary for interference are down by a loop factor and small
Cabibbo-Kobayashi-Maskawa (CKM) matrix elements, and there is no heavy
virtual top quark which could provide substantial breaking of the
Glashow-Iliopoulos-Maiani (GIM) mechanism. Naively, one would thus
expect effects of order $\mathcal{O}([V_{cb}V_{ub}/V_{cs}V_{us}]
\alpha_s/\pi) \sim 0.01\%$.

We define the amplitudes for final state $f$ as
\begin{equation}
\begin{split}
A_f & \equiv  A(D \to f ) = A^T_{f} \big[1+r_f
  e^{i(\delta_f-\phi_f)}\big],\\ 
\overline{A}_{f} & \equiv A(\bar D \to f ) =  A^T_{f}\big[1+r_f
  e^{i(\delta_f+\phi_f)}\big]\, .
\end{split}
\end{equation}
Here $A^T_{f}$ is the dominant tree amplitude and $r_f$ the relative
magnitude of the subleading amplitude, carrying the weak
phase $\phi_f$ and the strong phase
$\delta_f$. We can now define the direct $CP$ asymmetry as
\begin{equation}
\begin{split}
{\cal A}_f^{\rm dir} \equiv 
{|A_f|^2 -|\bar A_{f}|^2 \over |A_f|^2  + | \bar A_{ f } |^2 }  = 2
r_f \sin \gamma \sin \delta_f\, , 
\end{split}
\end{equation}
where the last equality holds up to corrections of
$\mathcal{O}(r_f^2)$. LHCb and CDF measure a time-integrated $CP$
asymmetry. The approximately universal contribution of indirect $CP$
violation cancels to good approximation in the difference
\begin{equation}
\begin{split}
\Delta {\cal A}_{CP}={\cal A}_{CP}(D\to K^+K^-) -
{\cal A}_{CP}(D\to \pi^+\pi^-)\, .
\end{split}
\end{equation}

The measurements of LHCb, $\Delta {\cal A}_{CP} = (-0.82 \pm 0.21 \pm
0.11)\%$~\cite{Aaij:2011in}, CDF, $\Delta {\cal A}_{CP}= (-0.62 \pm
0.21 \pm 0.10)\%$~\cite{CDF-talk}, and inclusion of the indirect $CP$
asymmetry $A_\Gamma$~\cite{Aubert:2007en,Staric:2007dt}, lead to the
new world average (including the Babar~\cite{Aubert:2007if},
Belle~\cite{Staric:2008rx}, and CDF~\cite{Aaltonen:2011se}
measurements) $\Delta {\cal A}_{CP}= (-0.67 \pm
0.16)\%$~\cite{CDF-talk}. In the following, we will try to answer
three questions: Can this measurement be accounted for by the standard
model? Can it be new physics? Can we distinguish the two
possibilities?


\section{Setting the stage}\label{sec:stage}

As a first step, we take the size of the tree amplitudes $A^T$ from
data and then relate the tree amplitudes to the penguin amplitudes
$A^P$ to estimate the size of the latter~\cite{Brod:2011re}. The
starting point of our analysis is the weak effective Hamiltonian
\begin{equation}\label{eq:Heff}
\begin{split}
H_{\rm eff}^{\rm SCS} = \frac{G_F}{\sqrt{2}} & \left\{
\left(V_{cs} V_{us}^*- V_{cd} V_{ud}^*\right)  \sum_{i=1,2} C_i \left(Q_i^{\bar s s} - Q_i^{\bar dd} \right)/2 \right. \\
&\left. -V_{cb} V_{ub}^* \,\left[ \sum_{i=1,2} C_i  \left(Q_i^{\bar s s} + Q_i^{\bar d d} \right)/2  +  \sum_{i=3}^{6} 
C_i Q_i + C_{8g} Q_{8 g} \,\,\right]\right\}
+ {\rm h.c.}\, . 
\end{split}
\end{equation}
The Wilson coefficients of the tree operators $Q_1^{\bar p p'} = (\bar
p u)_{V-A} \otimes (\bar c p')_{V-A}$, $Q_2^{\bar p p'} = (\bar
p_\alpha u_\beta)_{V-A} \otimes \\ (\bar c_\beta p'_\alpha)_{V-A}$,
the penguin operators $Q_{3\ldots 6}$, and the chromomagnetic operator
$Q_{8g}$, can be calculated in perturbation theory. The hadronic
matrix elements are harder to compute; we will estimate their size
using experimental data. They receive leading power contributions and
power corrections in $1/m_c$, which are expected to be large. 

A leading power estimation, using naive factorization and
$\mathcal{O}(\alpha_s)$ corrections, yields for the ratio
$r_f^\text{LP} \equiv |A^P_f(\text{leading
  power})/A^T_f(\text{experiment})|$: $r_{K^+K^-}^\text{LP} \approx
(0.01 - 0.02)\%$, $r_{\pi^+\pi^-}^\text{LP} \approx (0.015 -
0.028)\%$. This is consistent with, yet slightly larger than the naive
scaling estimate. We expect the signs of ${\cal
  A}_{K^+K^-}^\text{dir}$ and ${\cal A}_{\pi^+\pi^-}^\text{dir}$ to be
opposite, if $SU(3)$ breaking is not too large; so for $\phi_f =
\gamma \approx 67^\circ$ and $\mathcal{O}(1)$ strong phases we obtain
$\Delta {\cal A}_{CP} (\text{leading power}) = \mathcal{O}(0.1\%)$, an
order of magnitude smaller than the measurement.

However, we know from $SU(3)$
fits~\cite{Cheng:2010ry,Pirtskhalava:2011va,Cheng:2012wr,Bhattacharya:2012ah,Feldmann:2012js}
that power corrections can be large. To be specific, we look at
insertions of the penguin operators $Q_4$, $Q_6$ into power-suppressed
annihilation amplitudes. The associated penguin contractions of $Q_1$
cancel the scale and scheme dependence. Estimating their size using
the loop functions $G$, defined in~\cite{Beneke:2001ev}, and using
naive $N_c$ counting to relate the penguin to the tree amplitudes, we
arrive at $r_{f,1}^\text{PC} \approx (0.04 - 0.08)\%$,
$r_{f,2}^\text{PC} \approx (0.03 - 0.04)\%$, where $r_{f,i}^\text{PC}
\equiv |A^P_f(\text{power correction})/A^T_f(\text{experiment})|$ and
the subscripts $1,2$ correspond to the insertions of $Q_4$, $Q_6$,
respectively. Again assuming $\mathcal{O}(1)$ strong phases, this
leads to $\Delta {\cal A}_{CP} (r_{f,1}) = \mathcal{O}(0.3\%)$ and
$\Delta {\cal A}_{CP} (r_{f,2}) = \mathcal{O}(0.2\%)$ for the two
insertions. Thus, a standard model explanation seems plausible.

Of course, the extraction of the annihilation amplitudes from data,
neglected contributions to the annihilation amplitudes, $N_c$
counting, the modeling of the penguin contraction amplitudes, and the
neglected additional penguin contractions lead to an uncertainty of a
factor of a few. 
So, can we trust the estimate?

\section{A consistent picture}\label{sec:cp}

Another interesting observation is the large difference of SCS
branching ratios, $\text{Br}(D^0\to K^+K^-) \approx 2.8 \times
\text{Br}(D^0\to \pi^+ \pi^-)$. It implies that the ratio of
amplitudes (normalized to phase space) is $A(D^0\to K^+K^-) \approx
1.8 \times A(D^0\to \pi^+ \pi^-)$, whereas they would be equal in the
$SU(3)$ limit. This has often been interpreted as a sign of large
$SU(3)$ breaking. On the other hand, the ratio of the Cabibbo-favored
(CF) to the doubly Cabibbo-suppressed (DCS) amplitude is $A(D^0\to
K^-\pi^+) \approx 1.15 \times A(D^0\to K^+ \pi^-)$, after accounting
for CKM factors, in accordance with nominal $SU(3)$ breaking of
$\mathcal{O}(20\%)$.

A glance at the effective Hamiltonian~\eqref{eq:Heff} shows that the
combination $P$ of penguin contractions of $Q_{1,2}^{\bar s s} $ and
$Q_{1,2}^{\bar dd}$ proportional to $V_{cb} V_{ub}^*$ is $U$-spin
invariant, while $P_\text{break}$, the combination contributing to the
tree amplitude vanishes in the $U$-spin limit. $P_\text{break}$
contributes with opposite sign to the two SCS decay rates, and $P$
gives rise to a nonvanishing $\Delta {\cal A}_{CP}$. Guided by the
considerations exposed in Section~\ref{sec:stage}, we perform a
$U$-spin decomposition of the amplitudes to all four (CF, SCS, DCS)
decays, and fit these amplitudes to the data (branching ratios and
$CP$ asymmetries) under the additional assumption that penguin
contractions are large, of order $\mathcal{O}(1/\epsilon)$, where
$\epsilon \ll 1$.

\begin{figure}
\rule{5cm}{0.2mm}\hfill\rule{5cm}{0.2mm}\hfill\rule{5cm}{0.2mm}
\vskip 0.2cm
\includegraphics[scale=0.205]{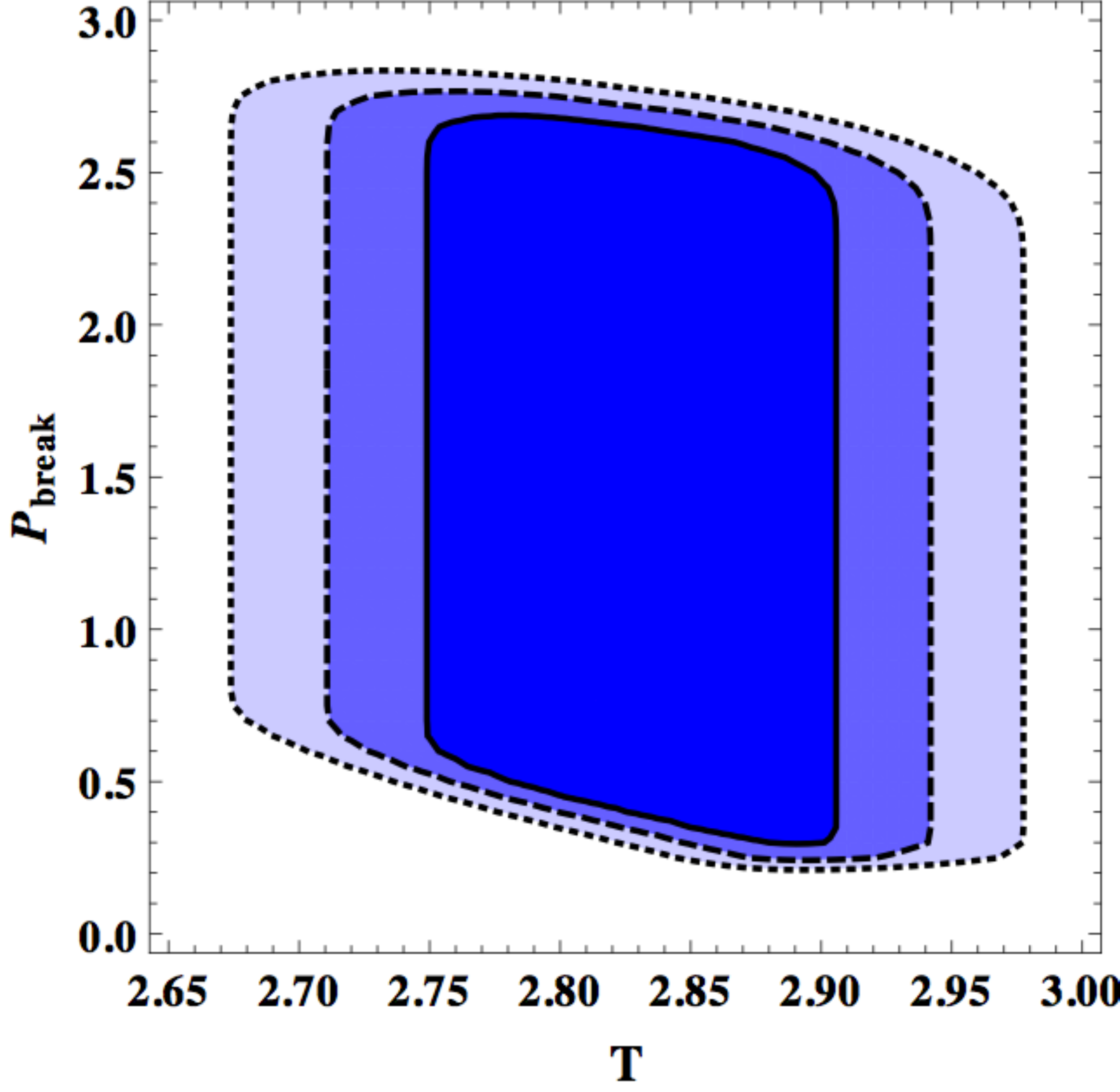}~~~~~~~~
\includegraphics[scale=0.17]{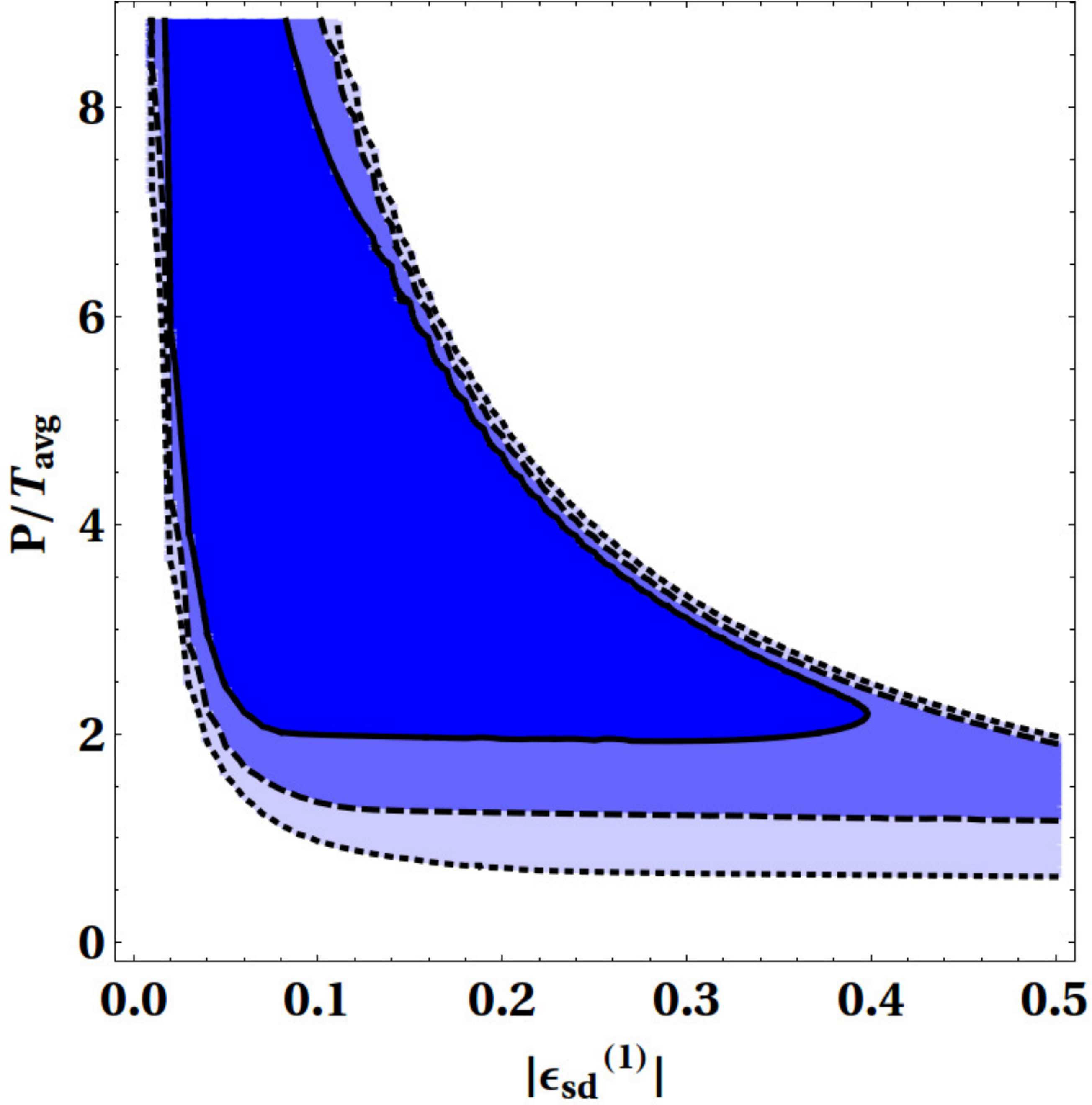}~~~~~~~~
\includegraphics[scale=0.18]{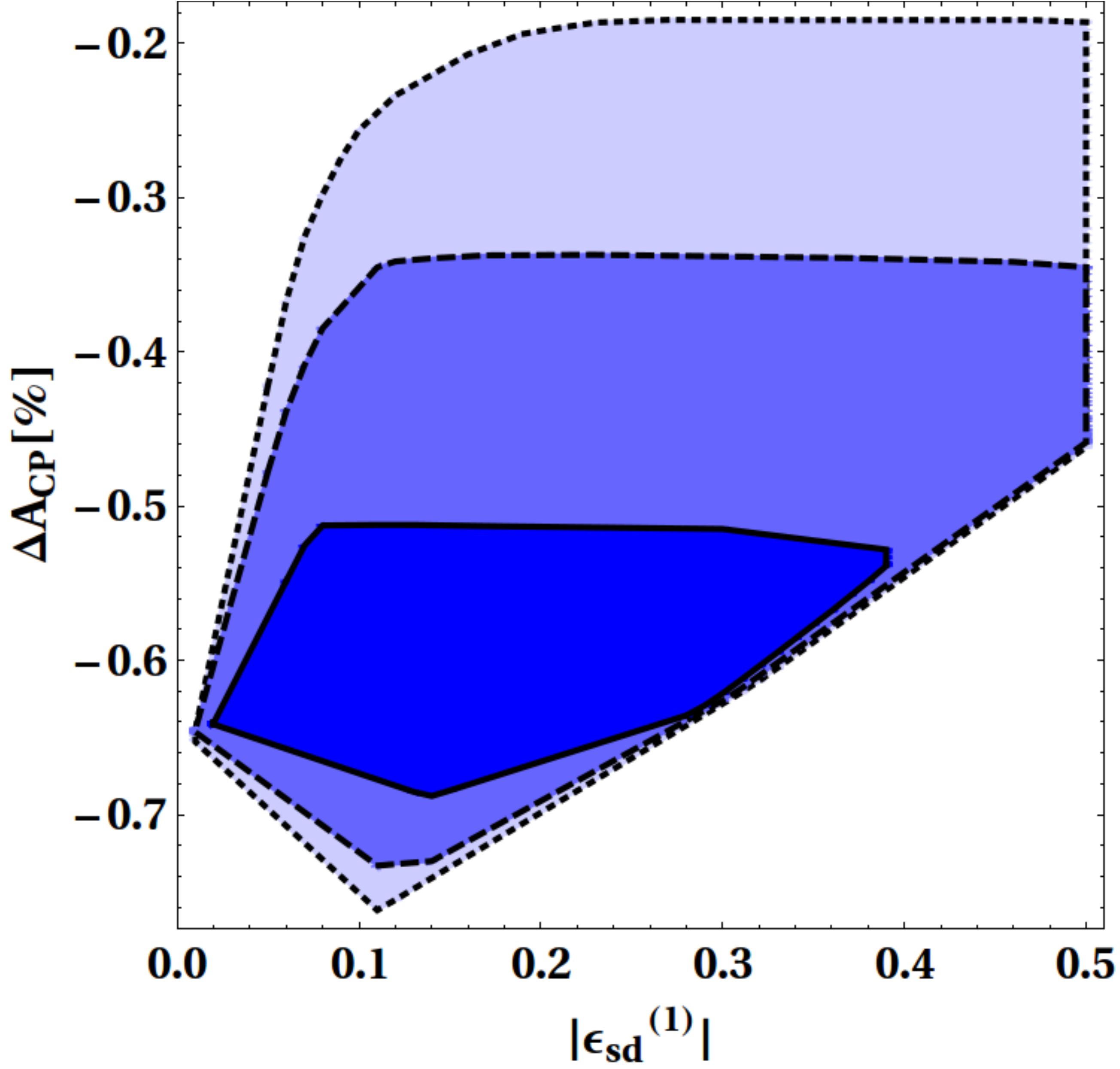}
\rule{5cm}{0.2mm}\hfill\rule{5cm}{0.2mm}\hfill\rule{5cm}{0.2mm}
\caption{The results of our fit. Solid, dashed, and dotted lines
  denote one-, two-, and three-sigma contours, respectively. Left
  panel: A fit to the branching ratios only yields $P_\text{break}
  \equiv \epsilon_{sd}^{(1)} P \sim T$, assuming nominal $U$-spin
  breaking. $T$ is the tree amplitude. The lower bound of
  $P/T_\text{avg}$ in the middle panel is directly related to the
  large difference of decay rates for the SCS modes. ($T_\text{avg}$
  is the average value of $T$ from the fit). It translates into the
  upper bound on $\Delta {\cal A}_{CP}$ -- the fit results can
  naturally accommodate the measured value (right panel).
\label{fig:1}}
\end{figure}

Our main point is~\cite{Brod:2012ud} that under the assumption of
nominal $U$-spin breaking, a broken penguin $P_{\rm break}$ which
explains the difference of the $D^0 \to K^+K^-$ and $D^0 \to
\pi^+\pi^-$ decay rates implies a $\Delta U=0$ penguin $P$ that
naturally\footnote{An important side remark is that no fine tuning of
  strong phases is required~\cite{Brod:2012ud}.} yields the observed
$\Delta {\mathcal A}_{CP}$.  The scaling $P_{\rm break } \sim
\epsilon_U P$ together with our fit result $P_{\rm break } \sim T/2$
(see Fig.~\ref{fig:1}) yields the estimate
\begin{equation} \label{randPovT} 
r_{\pi^+\pi^-, K^+K^-}\simeq
\left|\frac{V_{cb}V_{ub}}{V_{cs}V_{us}}\right|\cdot
\left|\frac{P}{T\pm P_{\rm break}}\right|\sim
\frac{|V_{cb}V_{ub}|}{|V_{cs}V_{us}|} \frac{1}{2\,\epsilon_U}\sim 0.2\%,
\end{equation}
for $\epsilon_U \sim 0.2$. This is consistent with the measured
$\Delta {\mathcal A}_{CP}$ for $\mathcal{O}(1)$ strong phases. Some
results of our fit are shown in Figure~\ref{fig:1}.

By the same reasoning, exchanging the spectator quark we expect direct
$CP$ asymmetries of the same order ($\approx0.5\%$) in the decay modes
$D^+ \to K^+ \overline{K^0}$, $D_s^+ \to \pi^+ K^0$.

\section{Can it be new physics?}\label{sec:np}

Whereas a standard-model explanation seems plausible, it is not
excluded that new physics contributes partly to $\Delta {\cal
  A}_{CP}$. Any new-physics explanation has to respect constraints
from other observables like $D$- and $K$-meson mixing, or direct
searches, but substantial contributions are still
possible~\cite{Isidori:2011qw,Altmannshofer:2012ur}. Can we
discriminate them from the standard-model contributions? 

Models of new physics that have $\Delta I = 3/2$ contributions could
be separated from the standard-model background (an example would be a
scalar color-singlet weak doublet~\cite{Hochberg:2011ru}). To see
this, note that the standard-model tree operators have both $\Delta I
= 1/2$ and $\Delta I = 3/2$ contributions, while the standard-model
penguin operators are pure $\Delta I = 1/2$ (apart from neglegible
electroweak contributions). For instance, the $I = 2$ final state in
$D^+ \to \pi^+ \pi^0$ cannot be reached by standard-model penguin
operators, so any observed direct $CP$ asymmetry in this decay would
be a clear signal of new physics. More sophisticated isospin sum rules
can be constructed~\cite{Grossman:2012eb}.

If new physics induces only $\Delta I = 1/2$ transitions, it seems
necessary to build explicit models and look for their collider
signatures. The most plausible models include chirally enhanced
chromomagnetic penguin
operators~\cite{Grossman:2006jg,Giudice:2012qq}.


\section{Conclusion}

Large penguin contractions in the standard model can naturally explain
both the large difference of decay rates in the $D^0\to K^+K^-$ and
$D^0\to \pi^+ \pi^-$ modes and the observed $\Delta {\cal
  A}_{CP}$. However, new-physics contributions are not
excluded. Viable models exist and can possibly tested.

\section*{Acknowledgments}
I thank Yuval Grossman, Alexander Kagan, and Jure Zupan for the
pleasant and fruitful collaboration, the organizers of ``Recontres de
Moriond'' for the invitation to this inspiring conference, Emmanuel
Stamou for proofreading the manuscript, and the NFS for travel
support. The work of J.~B. is supported by DOE grant FG02-84-ER40153.

\end{document}